\documentclass[sigconf, nonacm]{acmart}
\AtBeginDocument{%
  }

\begin{document}

%%
%% The "title" command has an optional parameter,
%% allowing the author to define a "short title" to be used in page headers.
\title{Revisiting “Cooler is Better”: ITD-Aware Per-CPU Thermal Optimization for Sustainable Data Center Operation}

%%
%% The "author" command and its associated commands are used to define
%% the authors and their affiliations.
%% Of note is the shared affiliation of the first two authors, and the
%% "authornote" and "authornotemark" commands
%% used to denote shared contribution to the research.

\author{Jason Crop}
\affiliation{%
 \institution{Colorado State University}
  \city{Fort Collins}
  \state{Colorado}
  \country{USA}
}
\email{Jason.Crop@colostate.edu}
\orcid{0009-0002-6148-6887}

\author{Hayden Moore}
\affiliation{%
  \institution{Colorado State University}
  \city{Fort Collins}
  \state{Colorado}
  \country{USA}
}
\email{Hayden.Moore@colostate.edu}
\orcid{0009-0005-7693-703X}

\author{Sudeep Pasricha}
\affiliation{%
  \institution{Colorado State University}
   \city{Fort Collins}
   \state{Colorado}
   \country{USA}
}
\email{Sudeep.Pasricha@colostate.edu}
\orcid{0000-0002-0846-0066}

%%
%% By default, the full list of authors will be used in the page
%% headers. Often, this list is too long, and will overlap
%% other information printed in the page headers. This command allows
%% the author to define a more concise list
%% of authors' names for this purpose.

%%
%% The abstract is a short summary of the work to be presented in the
%% article.
\begin{abstract}
 As data center energy demand approaches grid-level constraints, optimizing conventional server infrastructure is essential for sustainable growth. The long-standing assumption that “cooler is better”, i.e., lower CPU temperatures reduce power, does not fully hold for modern low-voltage CPUs, where inverse temperature dependence (ITD) drives higher supply voltages at lower temperatures. This creates a non-monotonic performance-per-watt curve where efficiency peaks at an intermediate thermal point. In this paper, for the first time, we empirically characterize ITD on production Intel Xeon CPUs and demonstrate that efficiency-optimal temperatures are CPU part-specific, and frequently higher than typical data center operating conditions. Measurements from commercial cloud data center platforms (Amazon, Equinix) reveal that approximately half of modern high-power CPUs operate about 10°C below their efficiency-optimal thermal point. By implementing ITD-aware thermal grouping of CPUs and inlet temperature adjustments, data center operators can optimize facility-level cooling and overall sustainability. Our case study shows that this approach can reduce total data center energy by 4–13\% without sacrificing performance or reliability.
\end{abstract}

%%
%% The code below is generated by the tool at http://dl.acm.org/ccs.cfm.
%% Please copy and paste the code instead of the example below.
%%

%%
%% Keywords. The author(s) should pick words that accurately describe
%% the work being presented. Separate the keywords with commas.
%% A "teaser" image appears between the author and affiliation
%% information and the body of the document, and typically spans the
%% page.

%%\received{15 May 2026}
%%\received[revised]{12 March 2009}
%%\received[accepted]{5 June 2009}

%%
%% This command processes the author and affiliation and title
%% information and builds the first part of the formatted document.
\maketitle

\section{Introduction}

While AI has become a primary driver of infrastructure growth, recent forecasts indicate that 73\% of data center power demand in 2027 will still be dominated by traditional cloud services and enterprise workloads \cite{accomando2025ai}. Within these data centers, CPUs remain the single largest contributor to power consumption in servers, despite the proliferation of GPUs and specialized AI accelerators \cite{wang2024designing}. As energy availability becomes the “hard ceiling” for data center expansion, reclaiming efficiency from these high-volume traditional server systems through per-CPU thermal optimization is no longer optional—it is a prerequisite for sustainable growth. Achieving this requires moving beyond uniform cooling policies that treat all CPUs within the same Stock Keeping Unit (SKU) as electrically identical. In practice, each CPU in a SKU exhibits unique voltage, leakage, and dynamic power characteristics that shape its energy efficiency profile. In this work, we show that every CPU possesses a distinct optimal thermal operating point that maximizes performance-per-watt.

Improving efficiency at scale requires understanding how modern CPUs consume power across operating temperatures. Although it is commonly assumed that lower operating temperatures reduce energy consumption \cite{Vog14}, semiconductor physics reveals a more nuanced reality. In advanced CMOS technologies operating at reduced supply voltages, transistors exhibit inverse temperature dependence (ITD)—a well-documented phenomenon in which carrier mobility and threshold voltage interactions cause transistor delay to increase at lower temperatures when operating near minimum voltage conditions \cite{sassone2012investigating}, \cite{cho2012characterization}. To preserve timing correctness under these conditions, CPUs must increase supply voltage as temperature decreases. Consequently, voltage is no longer independent of temperature in modern server CPUs.

However, while the device- and circuit-level physics of ITD are well established, there remains limited documentation and research on how modern CPU power management firmware handles these effects, and how that behavior translates into system-level energy efficiency. Our study focuses on Intel Xeon CPUs, which provide high-resolution telemetry—including core-level voltage, temperature, frequency, and power—accessible through Model-Specific Registers (MSRs) \cite{intelMSR}. Leveraging this
visibility, we empirically characterize the voltage–temperature relationship across multiple CPUs. We observe that manufacturers embed Digital Thermal Sensors (DTS) throughout the CPU die to support temperature-aware voltage regulation. Rather than relying on static voltage guardbands, the CPU dynamically adjusts supply voltage in real time to compensate for ITD-driven delay variations. This voltage–temperature coupling is illustrated in Figure 1, which shows the measured rise in core operating voltage required to sustain multiple frequency points as temperature decreases. The data represents aggregate Intel  Xeon 8480 CPU behavior using the mean core voltage and average temperature reported across all on-die DTS locations.

While this voltage adjustment preserves functional correctness, it fundamentally alters the CPUs power–temperature relationship. Dynamic power scales quadratically with voltage, whereas leakage power increases exponentially with temperature. At lower temperatures, ITD-induced voltage increases counteract the expected leakage reduction, producing a power tradeoff that is not immediately intuitive. When evaluated under fixed-frequency (iso-performance) conditions, this interaction manifests as a non-monotonic power curve (Figure 2 blue curve). Figure 2 depicts the SPEC CPU2017 povray workload \cite{specPovray} executed under iso-performance conditions across a broad thermal envelope on a single-socket 350W Intel Xeon 8480 system. The results demonstrate that although static power decreases with cooling, dynamic power rises due to ITD-driven voltage compensation. \textit{The combined effect produces a distinct minimum in total power, an efficiency-optimal temperature point (highlighted dark blue circle), beyond which further cooling becomes counterproductive}. Below this point, the increase in dynamic power outweighs the reduction in leakage, resulting in higher overall energy consumption. Importantly, because each CPU exhibits unique leakage and voltage characteristics, this optimal efficiency point is not universal but part-specific (in this paper the term “part” refers to a specific CPU within a SKU).

\begin{figure}
    \centering
    \includegraphics[width=0.9\linewidth]{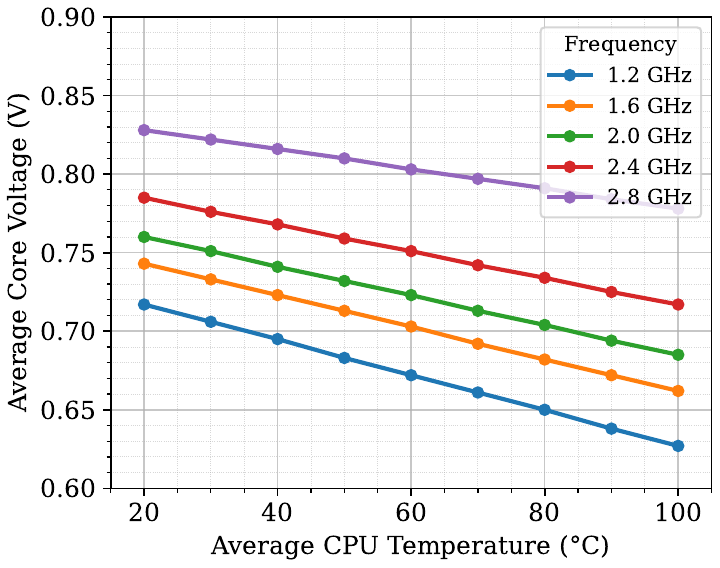}
    \vspace{-2mm} % This pulls the caption UP toward the image
    \caption{Measured core voltage versus temperature for Intel Xeon 8480 (Sapphire Rapids, 10nm) at various frequencies.}
    \vspace{-1mm} % Adjust this value as needed
\end{figure}
%%\FloatBarrier

\begin{figure}
    \centering
    \includegraphics[width=0.9\linewidth]{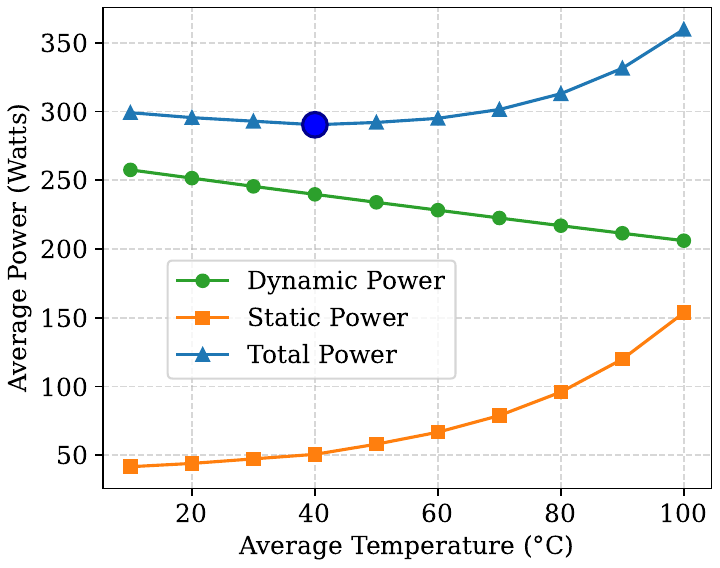}
    \vspace{-2mm} % This pulls the caption UP toward the image
    \caption{Xeon 8480 povray dynamic, static, and total power versus temperature at iso-performance (fixed frequency).}
    \vspace{-2mm} % Adjust this value as needed
\end{figure}

These findings challenge conventional thermal management assumptions in data centers where “cooler is better” is the norm and motivate system-level ITD-aware optimization. Rather than treating temperature reduction as an inherently beneficial strategy, we show why thermal control must be informed by the electrical behavior of modern CPUs with ITD power management. This shift in perspective—from uniform cooling to electrically informed temperature management—forms the foundation of our work. We make the following novel contributions in this paper:

\begin{itemize}
    \item We show that modern low-voltage, high-TDP server CPUs can exhibit a clear efficiency-optimal temperature, where cooling below that point reduces performance-per-watt.
    \item By accounting for ITD-driven voltage behavior in CPUs, we show how data center operators can safely shift thermal setpoints toward efficiency-optimal temperature points and reduce facility energy consumption.
    \item We further demonstrate that process-driven part-to-part variation shifts the optimal temperature point, motivating per-CPU characterization and rack-level thermal grouping rather than uniform policies in modern data centers.
\end{itemize}

\section{Related Work}

ITD has been studied in device and circuit contexts, where it has been shown that threshold-voltage shifts and carrier-mobility interactions can increase delay at lower temperatures under near-minimum-voltage operation \cite{sassone2012investigating}, \cite{cho2012characterization}. Prior ITD-focused research largely characterizes this behavior through transistor-level analysis, timing models, or selected circuit measurements, often with the goal of improving voltage margins or timing closure \cite{zu2016tistates}. In contrast, our work builds on these established physical principles to empirically quantify how ITD is handled by CPU firmware in modern server CPUs, and how it manifests as a system-level power–temperature tradeoff that can create an efficiency-optimal operating temperature. 

Numerous prior studies estimate CPU power through analytical models and performance event counters, with the primary goal of accurately predicting workload-dependent energy consumption \cite{sagi2020lightweight}, \cite{niaz2026concurrent}, \cite{vansteen2016analytical}. These approaches are effective for estimating dynamic power based on execution activity but typically treat the CPUs as electrically homogeneous within a SKU and assume temperature influences power primarily through leakage. Our work is complementary but focuses on a different question: how does temperature-dependent voltage regulation (driven by ITD) change dynamic power, and why does it produce non-monotonic power behavior at fixed performance? This is especially relevant for modern low-voltage, high-TDP CPUs, where voltage is actively coupled to temperature by design.

Building on these modeling foundations, recent work has expanded data center efficiency toward system- and fleet-level sustainability optimization, incorporating trade-offs across carbon emissions, water usage, and operational cost \cite{qi2024framework}, \cite{qi2026shield}. These approaches leverage geo-distribution to place AI and general workloads in regions with lower carbon intensity or reduced network-energy overheads \cite{hogade2025game}, \cite{hogade2022energy}, \cite{hogade2018minimizing}, while advanced resource management strategies account for thermal constraints and co-location interference to maintain throughput and avoid localized hotspots \cite{oxley2018rate}, \cite{oxley2016online}. Collectively, these frameworks provide effective control knobs at the data center scale, but typically rely on simplified or monotonic models of hardware power behavior. This work examines the CPU-level behavior underlying these abstractions, showing that voltage–temperature coupling can significantly influence dynamic power even at fixed performance, and highlighting the opportunity for more accurate hardware-level modeling to improve system-level sustainability strategies.

A substantial body of work addresses cooling efficiency \cite{cheong2019novel}, inlet temperature control \cite{vanle2024impacts}, and facility-level sustainability optimization \cite{shih2025research} in data centers. These studies frequently assume that cooler IT equipment reduces CPU power and thus emphasize aggressive cooling where feasible. Our results go against this assumption for modern CPUs: for parts operating near minimum voltage, ITD-driven voltage increases at low temperature can raise dynamic power enough to negate leakage reductions. This motivates temperature-aware, per-CPU thermal optimization as an important lever for sustainable operation, particularly in power-constrained environments where incremental efficiency gains translate directly to capacity and emissions benefits.

\section{ITD-Aware Modeling Framework}

To understand the non-monotonic behavior observed in Figure 2, we develop a lightweight ITD-aware modeling framework that captures how voltage and CPU core power evolve with temperature. While this work does not focus on a full detailed-level derivation, it is important to formalize the key electrical mechanisms that shape the system-level power curve.

Even when running the same workload, under identical operating system conditions, at fixed frequency (iso-performance) and fixed temperature (iso-temperature), measurable differences in voltage, leakage, and dynamic switching characteristics exist across CPU parts. Prior work by Kistowski et al \cite{kistowski2016variations} has demonstrated that CPUs within the same SKU can differ by up to 20\% in power consumption. Such variation is also associated with measurable performance differences across parts, as reported in our recent study \cite{jcrop2026var}. In our own measurements across ten CPU parts from the same SKU (Intel Xeon 6314 (IceLake)), we observed the following at a core frequency of 2.0GHz and average DTS of 50°C: 1) core voltage variation of up to 40 mV across CPU parts, 2) leakage differences approaching 2× between CPU parts, and 3) core dynamic capacitance (Cdyn) variation of up to 13\%. These variations directly influence the power–temperature curve and, critically, shift the efficiency-optimal temperature from one CPU to another. \textit{Thus, a critical premise of this framework is that CPUs within the same SKU are not electrically identical}.

\subsection{Lightweight ITD Voltage Model}

We first construct a simplified voltage model that captures the temperature-dependent behavior induced by ITD. As discussed earlier, ITD manifests when transistor delay increases at lower temperatures under near-minimum voltage operation, forcing the CPUs to raise supply voltage to maintain timing correctness. We define the following for a specific frequency:

\begin{itemize}
    \item $V_{MIN}$: voltage at 100°C with no ITD offset
    \item $V_{C}$: cutoff voltage above which ITD no longer exists
    \item $T_{C}$: cutoff temperature (100°C for Xeon CPUs)
    \item $T_{J}$: junction temperature (average of CPU)
    \item $ITD_{SLOPE}$: derived from measurements
\end{itemize}

Based on voltage–temperature measurements on real CPU chips (Figure 1), we model temperature-dependent core voltage as:
 \vspace{-0.5mm} % Adjust this value as needed
\begin{equation}
V = V_{MIN} + ((V_{C} - V_{MIN}) \times (T_{C} - T_{J}) \times ITD_{SLOPE})
\end{equation}

This expression captures the essential ITD behavior: as junction temperature decreases below the cutoff region, voltage must rise proportionally to maintain timing closure. Importantly, this model is not intended to be a detailed device-level formulation but rather a compact, system-oriented abstraction suitable for data center operation analysis and optimization studies.

\subsection{Simplified Power Model}

We next define a simplified CPU power model emphasizing core power, which dominates total consumption under high-load server operation.
 \vspace{-1mm} % Adjust this value as needed
\begin{equation}
P_{CPU} = ( C_{DYN} \times V^{2} \times F ) + ( I_{LEAK} \times V ) + P_{OTHER}
\end{equation}

Where $C_{DYN}$ is the workload-averaged effective switching capacitance of the cores, $V$ is the temperature-dependent core voltage from Eq. (1), $F$ is the fixed operating frequency (iso-performance), $I_{LEAK}$ represents leakage current at temperature $T_{J}$ and $P_{OTHER}$ captures the uncore and I/O baseline power. For clarity and tractability, we assume $P_{OTHER}$ is approximately constant across the thermal sweep. While we have measured temperature dependence in uncore power, its magnitude is secondary compared to core dynamic and leakage components under high power workloads.

To accurately capture the non-monotonic behavior of the power curve, the $I_{LEAK}$ term in Eq. (2) must account for the exponential relationship between leakage current and silicon temperature. While leakage is influenced by both voltage and temperature, for this lightweight framework, we focus on the dominant thermal effect. We define leakage current as an exponential function of junction temperature $T_{J}$:
 \vspace{-1mm} % Adjust this value as needed
\begin{equation}
I_{LEAK} = I_{0^{\circ}C} \times exp^{LTE \times T_{J}}
\end{equation}

Where $I_{0^{\circ}C}$ represents the base leakage current at 0°C and $LTE$ is the empirical temperature scaling factor.

This simplified formulation highlights two competing temperature-driven effects. First, dynamic power scales with $V^{2}$, meaning that any ITD-driven voltage increase at lower temperatures significantly raises switching energy. Second, leakage current increases exponentially with temperature, elevating static power at higher temperatures. Because voltage itself is a function of temperature (Figure 1), dynamic power can no longer be treated as temperature-independent. The interaction between these two effects fundamentally reshapes the total power curve and gives rise to the non-monotonic behavior observed in modern server CPUs.

\subsection{Parametric Sensitivity Temperature Shift}

Using this simplified ITD-aware power model, we performed parametric sweeps to evaluate how key electrical characteristics influence the efficiency-optimal temperature. Specifically, we varied leakage magnitude, effective $C_{DYN}$, and core voltage. The resulting trends are illustrated in Figure 3, where the highlighted points indicate the temperature corresponding to minimum total power under iso-performance operation (and also represent the operating point with highest performance-per-watt). Note that the baseline (blue curve) is the same total power curve in Figure 2.

\begin{figure}
    \centering
    \includegraphics[width=0.9\linewidth]{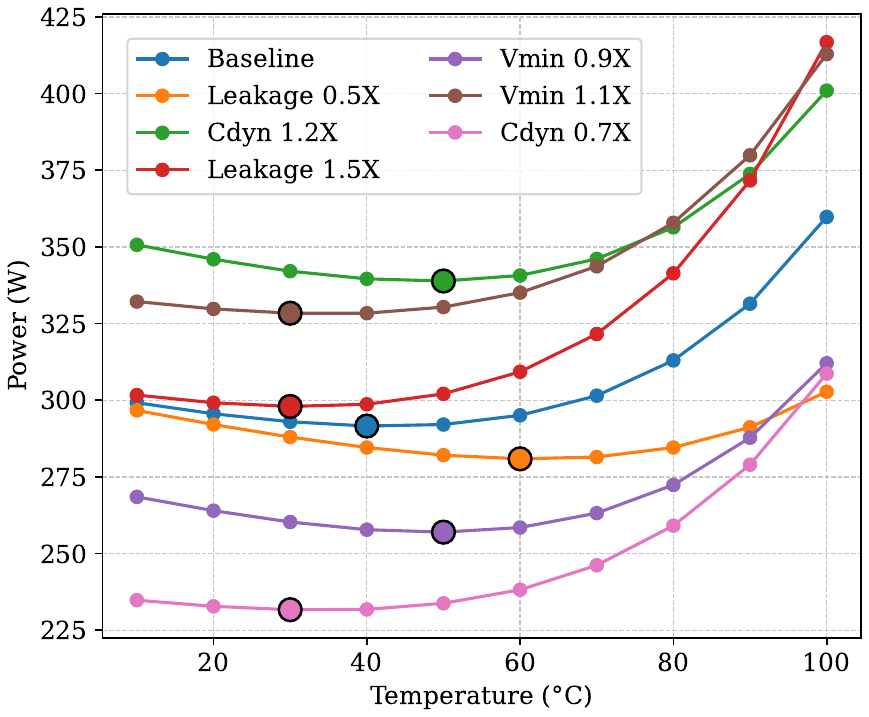}\
    \vspace{-2mm} % Adjust this value as needed
    \caption{Xeon 8480 \textit{povray} at Core=2.0 GHz model sweep of Leakage, CDYN, and voltage across temperature.}
    \vspace{-2mm} % Adjust this value as needed
\end{figure}

Three consistent behaviors emerge. First, lower leakage shifts the optimal temperature upward. As leakage is reduced, the high-temperature penalty diminishes. Second, higher dynamic activity (larger $C_{DYN}$) also raises the optimal temperature, as increased sensitivity to voltage amplifies the dynamic power associated with ITD-driven voltage. Third, lower voltage pushes the efficiency peak toward higher temperatures. Parts that operate closer to their minimum voltage are more sensitive to ITD, which increases the dynamic penalty of over-cooling.

Collectively, these relationships demonstrate that each CPU possesses a unique thermal efficiency profile dictated by its electrical parameters. Even within the same SKU, variation in leakage, dynamic capacitance, and minimum operating voltage produces distinct optimal temperature points. This part-specific behavior forms the foundation for fleet-level ITD-aware optimization discussed in the next section.

\section{Temperature-Aware Optimization Strategy}

Before proposing a system-level optimization strategy, it is first necessary to demonstrate that the efficiency-optimal temperature point is not limited to a single high power uniform workload ($povray$) but is consistently observable across diverse application behaviors. To evaluate robustness across workload classes, we executed the full SPEC CPU2017 benchmark suite on a modern Intel Xeon 8480 (56 cores, 350 W TDP) at a near-TDP operating point (2.0 GHz core, 1.6 GHz uncore). The suite includes both compute-intensive and memory-bound workloads, enabling evaluation under varied core and memory intensive distributions. The results, shown in Figure 4, reveal a consistent non-monotonic efficiency curve across workloads: each benchmark exhibits a distinct optimal temperature at which performance-per-watt peaks (indicated by the highlighted point). Cooling below this point reduces overall efficiency. Note that for clarity, we show a selected subset of floating-point benchmarks in Figure 4 to avoid clutter, but the full SPEC CPU2017 suite produced similar trends.

\begin{figure}
    \centering
    \includegraphics[width=0.9\linewidth]{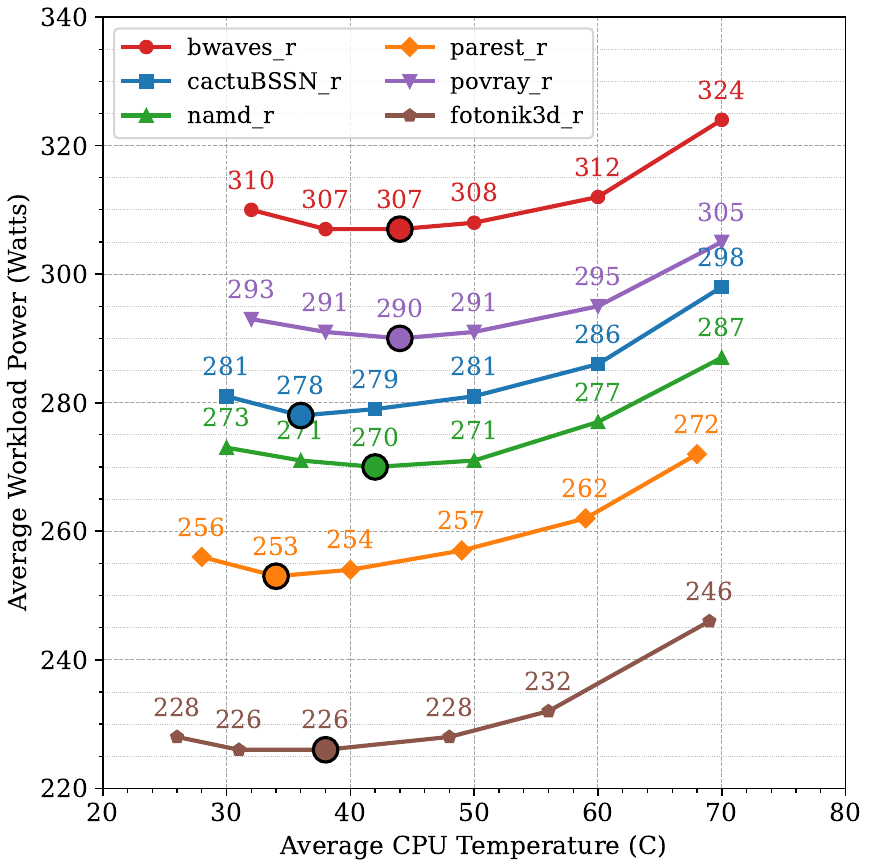}
    \vspace{-2mm} % Adjust this value as needed
    \caption{Xeon 8480 running selected SPEC CPU2017 benchmarks at iso-frequency at different temperatures.}
    \vspace{-2mm} % Adjust this value as needed
\end{figure}

During these thermal sweeps, operating frequency is held constant to ensure iso-performance conditions. Under such fixed performance, maximizing performance-per-watt is equivalent to identifying the temperature that minimizes total power consumption. This controlled methodology isolates electrical ITD and leakage effects from DVFS-induced performance variability.
Having established that the efficiency-optimal temperature point is observable across a broad range of workloads, several important questions emerge:

\begin{itemize}
    \item Why has this behavior not been widely documented in prior data center literature, and how is it expected to evolve in future server CPU generations?
    \item Are current data center cooling strategies aware of this optimal temperature region, or are many systems operating below their efficiency envelope due to over-cooling?
    \item How can data center operators practically leverage this CPU part-specific optimal temperature to improve fleet-level energy efficiency?
\end{itemize}

The following sections address these questions by examining architectural trends, evaluating current cooling practices, and proposing a scalable ITD-aware fleet optimization framework. 

\subsection{Why has this behavior gone undocumented?}

Examining historical trends in Intel Xeon server CPU architectures reveals two structural shifts that help explain why this efficiency-optimal temperature behavior has only recently become visible at the system level, as highlighted in this paper.

First, dynamic core power has increased dramatically across generations. A decade ago, Xeon server CPUs typically operated below 150 W TDP with fewer than 18 cores. Today, flagship server CPUs approach 400–500 W TDP and exceed 100 cores \cite{intelXeon6}. As shown in the parametric sensitivity analysis (Figure 3), increasing dynamic activity (higher $C_{DYN}$ and aggregate switching power) shifts the efficiency-optimal temperature upward. While an ITD-driven crossover point likely existed even in earlier generations, it would have occurred at extremely low temperatures—well outside normal data center operating ranges—and therefore remained practically invisible. As core counts and total dynamic power have grown, this crossover point has shifted to the standard thermal envelope of modern servers, making the non-monotonic behavior observable under typical conditions.

Second, nominal operating voltages have steadily declined due to process scaling and architectural optimization. Modern CPUs operate at lower voltages than prior generations. As operating voltage decreases, the magnitude of ITD-driven voltage compensation increases. This amplifies the dynamic power penalty associated with over-cooling and increases the optimal temperature upward. In combination, rising dynamic power density and reduced voltages have elevated the ITD crossover point into the range of conventional data center temperatures.

A third contributing factor is the limited visibility of ITD behavior in production firmware. Although ITD is well understood at the device and circuit levels, there is comparatively little published documentation detailing how server-class CPUs dynamically adjust voltage as a function of temperature. Without high-resolution telemetry and controlled iso-performance measurements, the resulting non-monotonic power behavior can easily be misattributed to workload variability or measurement noise rather than recognized as a systematic voltage–temperature interaction.

Together, these architectural and telemetry factors explain why this phenomenon has remained largely undocumented in prior data center literature. \textit{As process nodes continue to scale and operating voltage reduces, this optimal temperature point is expected to increase in future high-density server platforms.}

This naturally raises the next question: if efficiency-optimal temperatures now fall within normal operating ranges, are data centers currently operating at or below these points—and if not, what energy implication does that have?

\subsection{Are data centers over-cooling CPUs?}

To determine whether commercial data centers operate CPUs near their efficiency-optimal temperature, we conducted measurements on two cloud data center platforms that provide full bare-metal access to their servers: (i) Equinix Bare Metal Servers \cite{equinixMetal} and (ii) Amazon EC2 Bare Metal Servers \cite{amazonEC2}. Bare-metal access allowed complete control over workload execution and thermal characterization.

Our study included twenty CPUs across two Xeon generations: ten Gen3 32-core, 205 W Ice Lake CPUs hosted on Equinix systems \cite{intelXeon6314U}, and ten Gen4 48-core, 385 W Sapphire Rapids CPUs on Amazon EC2 \cite{intelXeon4thGen}. For each CPU, we measured per-part voltage, dynamic capacitance, and leakage parameters under native operating conditions. These measurements were used to construct individualized ITD-aware power models and estimate each part’s full power–temperature efficiency curve.

\begin{figure}
    \centering
    \includegraphics[width=0.9\linewidth]{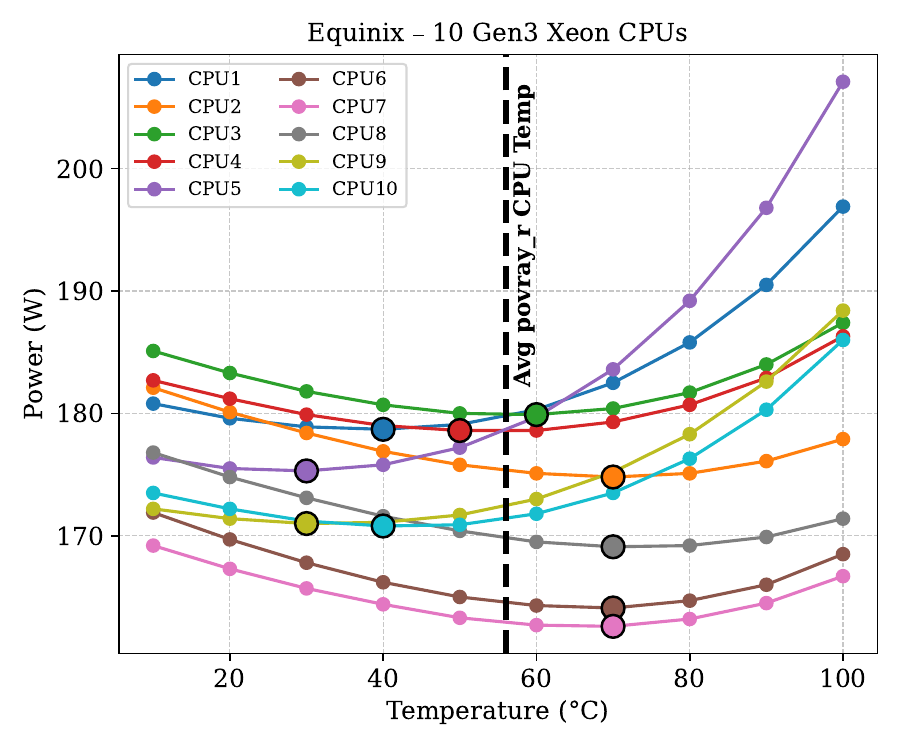}
    \vspace{-4mm} % Adjust this value as needed
    \caption*{(a)}

    \vspace{0.1cm}
    
    \includegraphics[width=0.9\linewidth]{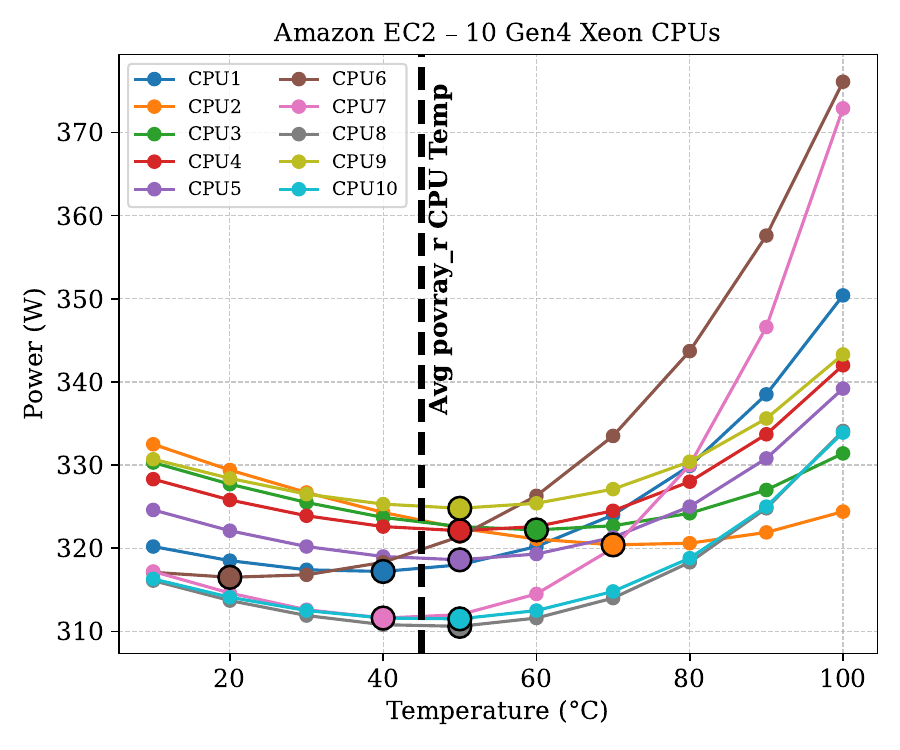}
    \vspace{-4mm} % Adjust this value as needed
    \caption*{(b)}
    \vspace{-4mm} % Adjust this value as needed
    \caption{Measured and modeled CPU power vs temperature for ten CPU parts per data center (a) Gen 3 Xeon (32-core, 205W TDP) on Equinix servers (b) Gen 4 Xeon (48-core, 385W TDP) on Amazon EC2 servers. Highlighted dots are optimal CPU part temperatures to achieve highest perf/W.}
\end{figure}

Figure 5(a) and 5(b) show the resulting efficiency curves for all measured parts in the Equinix and Amazon data centers. Each CPU exhibits a distinct optimal temperature point (highlighted), confirming that ITD-driven efficiency behavior is strongly CPU part-specific rather than uniform across a SKU. Even within identical CPU models, optimal temperatures ranged from approximately 20°C to 70°C, indicating notable electrical diversity.

When comparing modeled optimal temperatures to the actual operating temperatures observed in the cloud data center environments (shown as dashed vertical lines), we found that a substantial fraction of CPUs were operating below their efficiency-optimal point (indicated by CPUs with highlighted dots to the right of the vertical dashed line). Approximately half of the Gen3 Ice Lake CPUs on Equinix systems ran about 10°C below their optimal temperature. A similar trend appeared for the Gen4 Sapphire Rapids CPUs on Amazon EC2, where more than half of the parts also operated roughly 10°C below their efficiency peak.

These findings indicate that over-cooling is not isolated but recurring across data center platforms and CPU generations. Because performance was held constant, operating below the optimal temperature provides no throughput or reliability benefit; instead, it increases dynamic power due to ITD-driven voltage compensation while also raising facility cooling energy.

These results suggest that current data center cooling strategies do not account for part-specific efficiency-optimal temperatures. Uniform inlet or rack-level policies inevitably cause some CPUs to operate below their efficiency envelope, resulting in avoidable energy waste and motivating an ITD-aware optimization framework aligned with per-part electrical behavior.

\subsection{Strategies for energy efficiency}

If data center operators can characterize each CPU and determine its efficiency-optimal temperature, practical fleet-level optimization strategies become possible. A key question is which workload should define the optimal temperature setpoint. Our analysis of the SPEC CPU2017 suite shows that optimal temperatures across workloads generally fall within a narrow $\sim$10°C band (Figure 4). Although memory-bound workloads may exhibit slightly lower optimal temperatures, they draw substantially less power and impose less stress on cooling infrastructure. For this reason, optimization should be anchored to high-power, core-centric workloads that represent worst-case thermal conditions. The \textit{povray} benchmark from SPEC CPU2017 is an appropriate reference because it produces sustained, near-TDP, uniform core activity. Optimizing for this scenario ensures cooling infrastructure is dimensioned for peak demand while maximizing return on investment (ROI) in thermal management. Lower-power workloads naturally remain within efficient operating regions once the high-power case is optimized.

At the data center level, CPUs should be grouped according to their efficiency-optimal temperature. CPU parts with higher crossover temperatures can be co-located in racks or zones with elevated inlet temperatures, reducing cooling energy without sacrificing performance-per-watt. Conversely, CPU parts that benefit from lower temperatures can be grouped and cooled more aggressively. From a facility perspective, higher inlet temperatures improve cooling efficiency through an increased coefficient of performance (COP). In typical air-cooled data centers, chiller COP improves approximately 2--3\% per 1°C rise in supply air temperature \cite{globalEnergy2023}. Consequently, a controlled 10°C inlet increase can yield a 20--30\% improvement in cooling efficiency. When combined with IT-side reductions from avoiding ITD-driven over-cooling, this compounded effect produces meaningful facility-level energy savings, lowering both total cost of ownership (TCO) and carbon intensity per unit of compute.

To quantify these impacts, we extended a data center simulator \cite{moore2025sustainable} to incorporate our measured temperature-dependent CPU power and chiller COP behaviors. We modeled a 1,000-node data center facility executing five SPEC2017 benchmark classes; because inlet temperature also affects CPU power, the simulator integrates iso-efficiency curves derived from our measured performance-per-watt behavior (Figure 4). As illustrated in Figure 6, increasing the inlet temperature from 20°C to 30°C reduces energy by approximately 25\%, while a further increase from 30°C to 40°C provides an additional 8\% reduction. Because our measurements indicate that roughly half of the CPUs are operating below their efficiency-optimal temperature, only a subset of the fleet can safely realize these gains; accordingly, the net facility-level savings across the data center are estimated to be approximately 4–13\%, which is still quite substantial. Note that for clarity, we show only selected benchmarks in Figure 6 to avoid clutter, but the full benchmark suite produced similar trends.

Ultimately, ITD-aware thermal grouping enables data center operators to align electrical efficiency with cooling policy, transforming temperature from a fixed constraint into a dynamic optimization variable for sustainable data center operation.

\begin{figure}
    \centering
    \includegraphics[width=1\linewidth]{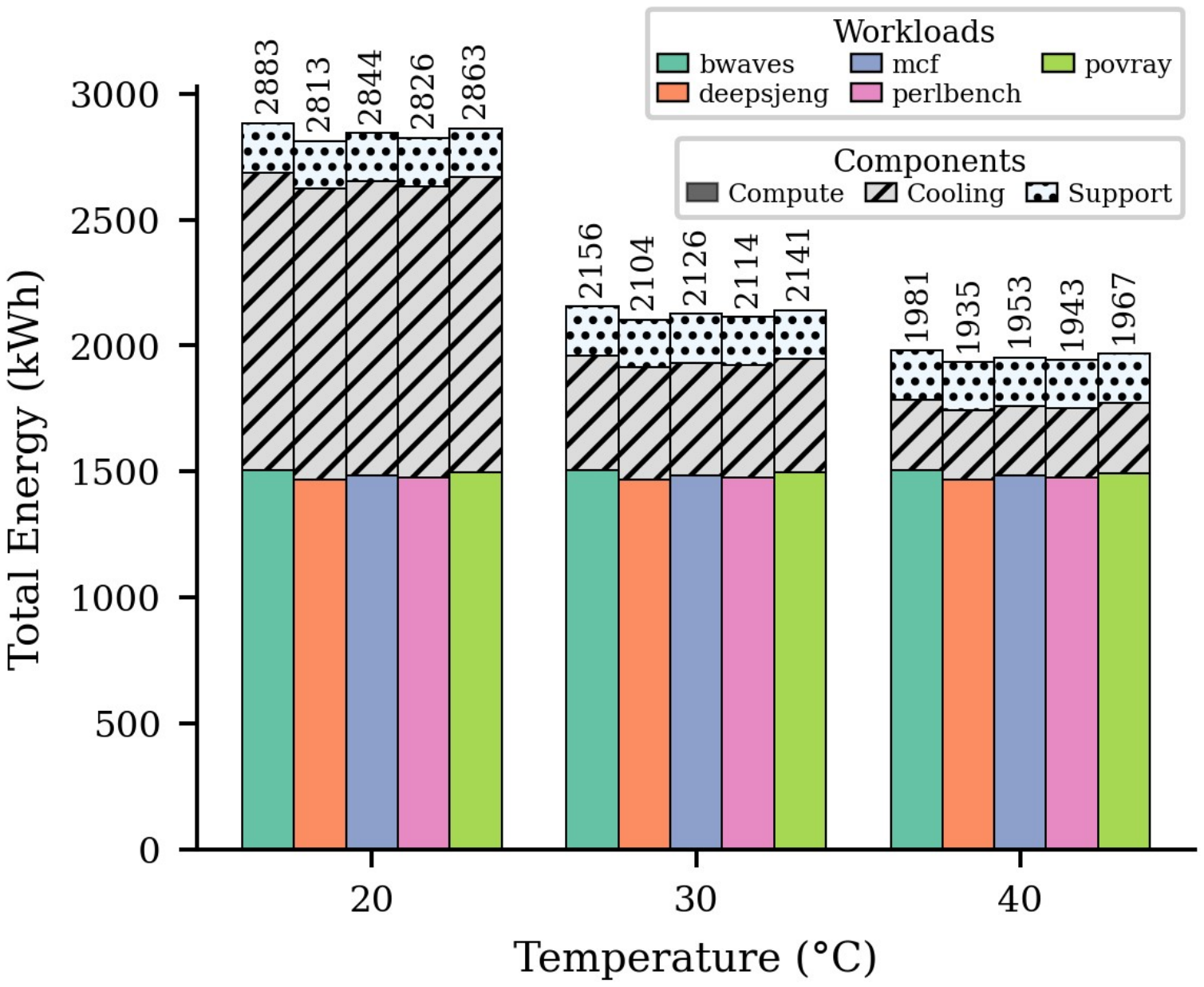}
    \vspace{-4mm} % Adjust this value as needed
    \caption{Total Energy vs Inlet Temp (Averages 20°C= 2846 kWh, 30°C=2128 kWh, 40°C=1957 kWh). Results are for a data center facility with 1000 Intel Xeon server nodes arranged in racks across aisles.}
    \vspace{-4mm} % Adjust this value as needed
\end{figure}

\section{Discussion: Higher Temperatures and Reliability}

Reliability is the primary concern when increasing operating temperatures. However, Xeon-class CPUs are architected for sustained operation up to approximately 100°C junction temperature, with thermal control policies typically transitioning to reliability-prioritized fan control near the defined $T_{CONTROL}$ threshold—often about 10°C below $T_{MAX}$ \cite{intelThermalGuide}. Raising average junction temperatures from 50–60°C to 60–70°C, as recommended in our work, remains within manufacturer-validated operating envelopes and preserves meaningful thermal margin. Operating within this range also ensures that long-term aging mechanisms are not materially accelerated, as the proposed setpoints remain comfortably below Intel’s thermal design limits. Many facilities operate with excess cooling headroom due to the long-standing assumption that “cooler is always better.” Our results indicate that increasing rack inlet temperatures by 10–20°C for selected processor groups can remain within reliability constraints while improving both IT efficiency and facility cooling performance.

\section{Conclusion}

 Modern low-voltage, high-TDP server CPUs no longer follow the traditional assumption that “cooler is always better.” Due to ITD, dynamic voltage increases at lower temperatures create a distinct efficiency-optimal operating point. Through empirical measurements across Equinix and Amazon data centers, and ITD-aware CPU power modeling, we demonstrate that this optimal temperature is part-specific and often higher than current data center operating conditions. ITD-aware thermal grouping and controlled inlet temperature increases can reduce cooling energy and deliver up to 13\% total data center facility-level energy savings, providing a practical pathway toward more sustainable, power-constrained data center operation without sacrificing performance or reliability. Our future work will characterize non-Intel server CPUs from AMD and ARM, as well as GPUs, to explore how our methodology can be more broadly applicable.

\begin{acks}
This research was supported in part by grants from HPE and the National Science Foundation (grants CCF-2450615, CCF-2324514, and CNS-2132385).

\end{acks}

%\vspace{0.5cm}
%\begin{minipage}{\linewidth}
%\nocite{*}
%\bibliographystyle{ACM-Reference-Format}
%\bibliographystyle{unsrtnat}
%\bibliography{refs}
%\end{minipage}

%%
%% The next two lines define the bibliography style to be used, and
%% the bibliography file.
\bibliographystyle{ACM-Reference-Format}
\bibliography{refs}

\end{document}